\documentclass[12pt]{article}
\usepackage{amsfonts}

\newfont{\twelvemsb}{msbm10 scaled\magstep1}
\newfont{\eightmsb}{msbm8}
\newfam\msbfam
\textfont\msbfam=\twelvemsb \scriptfont\msbfam=\eightmsb
\catcode`\@=11
\def\Bbb{\ifmmode\let\next\Bbb@\else
  \def\next{\errmessage{Use \string\Bbb\space only in math mode}}\fi\next}
\def\Bbb@#1{{\fam\msbfam{{#1}}}}

\newcommand{\be}{\begin{equation}}
\newcommand{\ee}{\end{equation}}
\newcommand{\ba}{\begin{eqnarray}}
\newcommand{\ea}{\end{eqnarray}}
\newcommand{\spz}{\hspace{0.5cm}}

\def\a{\alpha}
\def\d{\delta}
\newcommand{\D}{\Delta}

\newcommand{\p}{\partial}
\newcommand{\dx}{\partial_x}

\newcommand{\bZ}{{\Bbb Z}}

\newcommand{\NP}[1]{Nucl.\ Phys.\ {\bf #1}}
\newcommand{\PL}[1]{Phys.\ Lett.\ {\bf #1}}

\newcommand{\CMP}[1]{Comm.\ Math.\ Phys.\ {\bf #1}}
\newcommand{\CPAM}[1]{Comm.\ Pure\ Appl.\ Math.\ {\bf #1}}

\newcommand{\LMP}[1]{Lett.\ Math.\ Phys.\ {\bf #1}}

\begin{document}
\sloppy
\renewcommand{\thefootnote}{\fnsymbol{footnote}}

\newpage
\setcounter{page}{1}

\vspace{0.7cm}
\begin{flushright}
APCTP 2004-001\\
\end{flushright}
\vspace*{1cm}
\begin{center}
{\bf Geometrical Loci and CFTs via the Virasoro Symmetry of the mKdV-SG hierarchy: an excursus}\\
\vspace{1.8cm} {\large Davide Fioravanti  \footnote{E-mail:
df14@york.ac.uk}}\\
\vspace{.5cm} {\em Department of Mathematics, University of York, Heslington, York YO10 5DD, UK} \\
\end{center}
\vspace{1cm}

\renewcommand{\thefootnote}{\arabic{footnote}}
\setcounter{footnote}{0}

\begin{abstract}
{\noindent We} will describe the appearance of specific algebraic
KdV potentials as a consequence of a requirement on a
integro-differential expression. This expression belongs to a
class generated by means of Virasoro vector fields acting on the
KdV field. The ``almost'' rational KdV fields are described in
terms of a geometrical locus of complex points. A class of
solutions of this locus has recently appeared as a description of
any conformal Verma module without degeneration.

\end{abstract}

\vspace{1cm} {\noindent PACS}: 11.30-j; 02.40.-k; 03.50.-z


\newpage

\section {Introductory remarks}

A large well-known class of (abelian) isospectral deformations of
the Schr\"odinger operator $L=-\dx^2+u$ takes the name of
Korteweg-deVries ({\bf KdV}) hierarchy \cite{CDD}; for the
Korteweg-deVries equation
\begin{equation}
\frac{\p u}{\p t_1}=-u_{xxx}+6u_xu \label{KdV}
\end{equation}
is the first non-trivial example of flow in this hierarchy ($t_1$
and $x$ are time and space respectively, their appearance as
indeces means derivation). The entire KdV hierarchy can be derived
from the Schr\"odinger operator as the only fundamental object
\cite{CDD}, but it is also generated effectively by the
pseudo-differential operator
\begin{equation}
N_u=-\dx^2+4u+2u_x\dx^{-1}
\end{equation}
after successive actions on the trivial vector field $K_0[u]=u_x$
\begin{equation}
K_{j+1}=N_u K_j, \spz j\ge 0.
\end{equation}
And the actions of the different vector fields $K_j$ must be
understood in the usual way, as given by the time derivatives
\begin{equation}
\frac{\p u}{\p t_j}= K_j[u] . \label{hKdV}
\end{equation}
These flows do not change the spectrum of the Schr\"odinger
operator $L=-\dx^2+u$ \cite{CDD} and are all compatible, in the
sense that they commute with each other:
\begin{equation}
[K_i, K_j]=0. \label{comm}
\end{equation}
In this note we will not be self-contained and would like to
address the attentive reader to the main references \cite{ZM} and
\cite{DG} for definitions and theorems.

\section {The Virasoro vector fields and their decaying rational solutions}
In the same spirit, Zubelli and Magri \cite{ZM} constructed
recursively an algebra of vector fields \footnote{These vector
fields may also be derived from those of \cite{OS} putting
formally all the KdV times to zero.} (also called, in another
context, master-symmetries after Fokas and Fuchsteiner \cite{FF})
starting from the generator of the scaling transformation
\begin{equation}
V_0[u]=u+\frac{1}{2}xu_x \spz , \spz V_{j+1}=N_u V_j, \spz j\ge 0
. \label{vird}
\end{equation}
Albeit the flows of the KdV hierarchy commute with each other,
the flows $V_j$ were proved to close half centerless Virasoro
algebra (only generators for $j\geq 0$ are present):
\begin{equation}
[V_i, V_j]=(j-i)V_{i+j}. \label{vir}
\end{equation}
Moreover, the commutator with a higher KdV flow gives another KdV
vector field according to
\begin{equation}
[V_i, K_j]=(j+\frac{1}{2})K_{j+i}.
\end{equation}
Since the right hand side is not zero, the Virasoro vectors are
not proper symmetries of the KdV hierarchy, but they nevertheless
generate all the hierarchy by successive commutators starting only
from the first flow (\ref{KdV}). In contrast with the KdV flows,
they also change the spectrum of the Schr\"odinger operator,
though their action on the ``energy levels'' is simply realised by
the polynomial vector fields in the complex plane \cite{GO}.
Later on, how they act on the modified KdV field, $\phi$ such that
\begin{equation}
u=\phi_x^2+\phi_{xx},  \label{mKdV}
\end{equation}
has been studied in \cite{FS2}, showing that the algebra
(\ref{vird}) can be completed to form an entire Virasoro
algebra.\footnote{Actually, we should apologise because at that
time we did not know the content of \cite{ZM}. Later, we also had
a nice conversation with F. Magri about the algebraic structure of
\cite{FS2}, though we have discovered the content of \cite{ZM}
only very recently.} Furthermore, the Virasoro algebra has been
given a geometrical origin and meaning, by which it has been also
proved to enjoy a zero-curvature form (involving the differential
first order matrix operator associated to the second order
Schr\"odinger operator through the Miura transformation)
\cite{FS2} and to commute (genuine symmetry) with the
(light-cone) sine-Gordon flows \cite{FS3}. Going back to the
paper by Zubelli and Magri \cite{ZM}, they proved that the half
Virasoro algebra (\ref{vird}) is tangent to some previously
well-known manifolds: these are spanned by the bispectral
potentials $u(x)$ of the Schr\"odinger operator $L=-\dx^2+u$
\cite{DG}. In fact, Duistermaat and Gr\"unbaum characterised these
potentials to be the family of the pure angular momentum
potentials $u(x)=\frac{l(l+1)}{x^2}$, with $l(l+1)$ an arbitrary
constant, and two other classes: both are obtained by iterative
actions of {\it rational} Darboux transformations starting from
two very simple potentials, i.e. $u(x)=0$ and
$u(x)=-\frac{1}{4x^2}$. Therefore, they concluded that all the
bispectral potentials are {\it rational functions} decaying at
$x=\infty$ \footnote{To be rigorous, they also found the only
exception to zero-boundedness, i.e. the linear potential
$u(x)=c_1x+c_0$. But it was excluded from the analysis in
\cite{ZM} because of its triviality.} (with a peculiar pattern of
poles, cfr. \cite{DG} and below). As a consequence of tangency it
comes out that the (bounded) bispectral potentials stay rational
while they evolve according to one of the Virasoro vector fields
(\ref{vird}) \cite{ZM}. Recently, Zubelli and Silva have shown
the reverse statement: if $u(x)$ is a rational function decaying
at infinity, which remains rational by each of the flows
(\ref{vird}), then $u(x)$ is a (bounded) bispectral potential
\cite{ZS}. To prove it, they have almost used only the first
vector field obtained from the scaling transformation $V_0[u]$,
i.e.
\begin{equation}
V_1[u]=-\frac{x}{2}(u_{xxx}-6uu_x)-2u_{xx}+4u^2+u_x\dx^{-1}u .
\end{equation}
In fact, they have sought for decaying rational solutions,
decomposed in partial fractions
\begin{equation}
u_r(x)=\sum_{p\in P}\sum_{m=1}^{m_p}\frac{c_{p,m}}{(x-p)^m},
\label{rat}
\end{equation}
of the ``equation of motion''
\begin{equation}
\frac{\p u_r}{\p \beta_1}=V_1[u_r], \label{em1}
\end{equation}
where $\beta_1$ is the ``time'' associated with the flow under
examination and the poles $p=p(\beta_1)$ and the coefficients
$c_{p,m}=c_{p,m}(\beta_1)$ depend smoothly on it. Here they have
assumed that the number of poles is constant (neither poles
creation nor annihilation) and that the leading order coefficient
$c_{p,m_p}(\beta_1)$ must not vanish. They have discovered that
the ``generic'' rational function of the form (\ref{rat}), under
these conditions, undergoes severe restrictions on its form as
necessary conditions to satisfy the previous equation of motion
(\ref{em1}) and takes the simpler double pole form
\begin{equation}
u_r(x)=\frac{l(l+1)}{x^2} + \sum_{p\in P} \frac{2}{(x-p)^2},
\label{dpf}
\end{equation}
where $l(l+1)$ is a constant, that may also be zero, and each
$p=p(\beta_1)$ is a function of time. If $l(l+1)=0$ then $x=0$ is
not a stationary pole and then it may belong to the set of
complex non-stationary poles $P$; instead if $l(l+1)\neq0$ then
$x=0$ is a stationary pole and it does not belong to $P$. Now,
the time evolution of $u_r$ (\ref{dpf}) comes out by equating the
$1/(x-p)^3$ terms of both members of (\ref{em1}) and involves
only the poles $p(\beta_1)$
\begin{equation}
\dot{p}=-2(\frac{l(l+1)}{p}+\sum_{q\in P_p}\frac{2p+q}{(p-q)^2}),
\label{em1p}
\end{equation}
where the upper dot indicates the derivative with respect to the
time $\beta_1$ and we have for short defined $P_p=P-\{p\}$.
Finally, the poles in $P$ must satisfy a static constraint which
derives from equating to zero the $1/(x-p)^2$ terms of the right
hand side of (\ref{em1}):
\begin{equation}
\frac{l(l+1)}{p^3}+\sum_{q\in P_p} \frac{2}{(p-q)^3} =0.
\label{dglocus}
\end{equation}
When $l\in \bZ_{\geq0}$ or $l\in \bZ_{\geq0}-\frac{1}{2}$ these
relations give the celebrated {\it locus} of Duistermaat and
Gr\"unbaum \cite{DG}. As particular case, it reduces to the even
more famous locus of Airault, McKean and Moser if $l(l+1)=0$
\cite{AMM}
\begin{equation}
\sum_{q\in P_p} \frac{1}{(p-q)^3} =0 ,  \label{ammlocus}
\end{equation}
where now $P$ may contain the zero. The $u_r(x)$ satisfying the
previous condition are again given by the form (\ref{dpf}) with
now $l(l+1)=0$ and they are also the only rational solutions of
the KdV flows \cite{AMM} (in which case the poles would depend on
the KdV times $t_j$ (\ref{hKdV})). Moreover, they were obtained in
\cite{DG} by rational Darboux transformations from the initial
potential $u_0=0$. Instead, if $l(l+1)\neq0$ the potentials
(\ref{dpf}), (\ref{dglocus}) cannot be solutions of the KdV
hierarchy, albeit they are yielded by rational Darboux
transformations from $u_0(x)=\frac{-1/4}{x^2}$. Actually, the
locus (\ref{dglocus}) seems to be more general than in the
Duistermaat and Gr\"unbaum context since no restriction on $l$
comes out manifestly. In fact, this appears as a consequence of
imposing that the rational potentials are also solutions of the
higher Virasoro equations \cite{ZS}
\begin{equation}
\frac{\p u_r}{\p \beta_n}=V_n[u_r], \spz n>1. \label{emn}
\end{equation}
Actually, the poles of $u_r(x)$ cannot be considered proper
functions of all the times $\beta_j$ ($j>0$) at once, since the
different flows $V_j$ do not commute. Nevertheless, the
imposition of the rational invariance under the higher flows is
geometrically meaningful and suggestive. But it does not easily
imply that the restriction on $l$ might be seen as a consequence
of the only equation of motion (\ref{em1}), although it has been
conjectured in \cite{ZS}. Of course, there is no restriction on
$l$, if the only pole in $u_r(x)$ (\ref{dpf}) is $x=0$ (bispectral
potential $u_r(x)=l(l+1)/x^2$, which is scaling invariant:
$V_0[u_r]=0$).

\section {Rational potentials perturbed by $x^{2\alpha}$: CFTs}
\setcounter{equation}{0}

So far, we have been dealing with rational functions $u_r(x)$
(\ref{rat}) decaying at infinity and now we would like to consider
potentials which play some r\^oles in 1-dimensional quantum
mechanics. Moreover, as the whole Virasoro algebra was derived in
\cite{FS2} by dressing the infinitesimal generators of
diffeomorphisms in the spectral parameter, we expect it to have an
interesting action on more general potential. Therefore, we
consider a simple {\it perturbation} of (\ref{rat})
\begin{equation}
u(x)=x^{2\alpha}+u_r(x) \label{rat'}
\end{equation}
where $2\alpha$ is a non-negative real number. This anharmonic
perturbation \footnote{A coupling constant in front of
$x^{2\alpha}$ can be considered as re-absorbed through a
redefinition of $x$.} clearly changes the behaviour at $x=\infty$,
dominating the asymptotic expansion, and destroys any chance that
$u$ may be consistently a solution of equation (\ref{em1}).
Nevertheless, thanks to their aforementioned origin in \cite{FS2},
we may think of the vector fields (\ref{vird}) as infinitesimal
transformations acting on (\ref{rat'})
\begin{equation}
\d_j u=\epsilon_j V_j[u], \spz n\geq0, \label{infrat}
\end{equation}
where $\epsilon_j$ is an infinitesimal variation of $\beta_j$. In
this perspective, it is natural to require that these
transformations do not introduce new double poles; for this is the
requirement which gives rise to (\ref{dglocus}). Besides, a
(rational) Darboux transformation maps the solutions of the the
algebraic equations (\ref{dglocus}) into the solutions of a set of
equations with the same form (but different $l$)
\cite{DG}.\footnote{We do not expect that the Darboux
transformation will play exactly the same r\^ole here, since the
addition (\ref{rat'}) itself breaks the rationality.} In the first
instance, we do not want double poles in the first non-trivial
transformation of (\ref{rat'})
\begin{equation}
\d_1 u= V_1[u], \label{infrat1}
\end{equation}
where we have omitted $\epsilon_1$. Hence, if we limit
$2\alpha=r/s$ to be rational, we realise rather easily that we
need $u_r(x)$ in (\ref{rat'}) to be more stringently a rational
function of $y=x^{\frac{1}{s}}$, in order to equate the squared
powers of the transformation (\ref{infrat1}) to zero . And this
easily means that we can assume, without loss of generality,
$2\alpha$ to be a non-negative integer number. A similar but more
refined balancing of partial fraction decomposition yields, as
necessary condition, the restriction (\ref{dpf}) and hence
\begin{equation}
u(x)=x^{2\alpha}+\frac{l(l+1)}{x^2} + \sum_{p\in P}
\frac{2}{(x-p)^2}. \label{dpf'}
\end{equation}
Eventually, plugging again this form into the right hand side of
(\ref{infrat1}) we equate to zero the coefficients of the powers
$1/(x-p)^2$, obtaining $\forall p\in P$
\begin{equation}
l(l+1)+\sum_{q\in P_p} \frac{2p^3}{(p-q)^3} -\a p^{2\a+2} =0.
\label{dglocus'}
\end{equation}
These constraints can be thought of as a {\it deformation} of the
Duistermaat and Gr\"unbaum locus (\ref{dglocus}), in the sense
that they reduce to that locus when $\a=0$. Actually, we must
emphasise that so far no restriction on the possible values of $l$
has appeared. Nevertheless, we might think to deduce a restriction
by using the higher Virasoro vector fields, likewise to what
happens in the decaying potential case \cite{ZS}. This possible
restriction would be of crucial interest for what we are going to
illustrate. Moreover, we would like to deliver a more detailed
analysis of the geometrical interpretation of (\ref{dglocus'}) in
a future publication \cite{35}, where we also should shed light on
the creation and annihilation properties of the negative and
positive Virasoro generators of \cite{FS2}, respectively.

To gain some meaning from the algebraic equations
(\ref{dglocus'}), we will look for particular solutions. Thanks to
the presence of the last term in the left hand side, it would be
very natural to have solutions $p$ which gather in $(2\a+2)$-th
roots of another variable $z$. More precisely, we want that the
whole set $P$ is generated from another non-empty set of complex
numbers, $Z$, in this way:
\begin{equation}
P=\{p: p^{2\a+2}=z, \hspace{0.2cm} z\in Z\}. \label{Proots}
\end{equation}
Naturally, the sum in (\ref{dglocus'}) splits into two parts: $1)$
the sum over the $q$ which are not roots of $z=p^{2\a+2}$, i.e.
over $P_p^{(1)}=\{q\in P_p: q^{2\a+2}\neq p^{2\a+2}\}$; $2)$ the
sum over the roots $q$ of $z=p^{2\a+2}$ different from $p$, i.e.
over $P_p^{(2)}=\{q\in P_p: q^{2\a+2}=p^{2\a+2}\}$. As for the
case $1)$ we easily have
\begin{equation}
\sum_{q\in P_p^{(1)}} \frac{2}{(p-q)^3}=\frac{\p^3}{\p
p^3}\ln\prod_{q\in P_p^{(1)}} (p-q)= \frac{\p^3}{\p p^3}\sum_{w\in
Z_z} \ln(p^{2\a+2}-w),
\end{equation}
where again we have defined $Z_z=Z-\{z\}$. Therefore, after three
derivatives we obtain
\begin{equation}
\sum_{q\in P_p^{(1)}}
\frac{2}{(p-q)^3}=\frac{2\a+2}{p^3}\sum_{w\in Z_z}
\frac{2z[z^2+(1+2\a)(\a+3)wz+(1+2\a)w^2]}{(z-w)^3} . \label{s1}
\end{equation}
On the other hand the sum $2)$ over all the roots of $z=p^{2\a+2}$
except $p$ itself yields, after simple trigonometric
manipulations,
\begin{equation}
\sum_{q\in P_p^{(2)}} \frac{2}{(p-q)^3}=\frac{2\a+1}{2p^3}
 -\frac{3}{4p^3}\sum_{k=1}^{2\a+1}\frac{1}{\sin^2\frac{\pi}{2\a+2}k} .
\end{equation}
Moreover, the last sum can be computed explicitly (cfr. e.g.
\cite{GR}) with the simple outcome
\begin{equation}
\sum_{q\in P_p^{(2)}} \frac{2}{(p-q)^3}=\frac{1-4\a^2}{4p^3}.
\label{s2}
\end{equation}
Eventually, we collect both contributions (\ref{s1}) and
(\ref{s2}) into the locus equations (\ref{dglocus'})
\begin{equation}
\frac{4l(l+1)-4\a^2+1}{16(\a+1)}+ \sum_{w\in Z_z}
\frac{z[z^2+(1+2\a)(\a+3)wz+(1+2\a)w^2]}{(z-w)^3}
-\frac{\a}{4(\a+1)} z=0 , \label{blzlocus}
\end{equation}
which are the constraints $\forall z\in Z$. In the end, we remark
that this locus of complex points has been recently proposed by
Bazhanov, Lukyanov and A. Zamolodchikov \cite{BLZ} as describing
the non-degenerate conformal Verma module of highest weight
$\D=\frac{4l(l+1)-4\a^2+1}{16(\a+1)}$ with central charge
$c=1-\frac{6\a^2}{\a+1}$, where the cardinality of $Z$ gives the
level. This sort of description comes out ``naturally'' within a
surprising correspondence between the whole spectrum of a specific
Schr\"odinger operator $L=-\dx^2+u$ and the eigenvalue of the
Baxter Q-operator \cite{B} on a conformal state. As for the vacuum
the authors of \cite{DTBLZ} have first furnished the potential
(without the non-zero poles $p\in P$)
$u(x)=x^{2\alpha}+\frac{l(l+1)}{x^2}$. Therefore, the locus
(\ref{blzlocus}) extends the correspondence to the excited states
by means of exactly the rational potentials (\ref{dpf'}),
(\ref{Proots}) we have found. In addition, we have met a
restriction on $l$ in the previous case where we have obtained
(\ref{dglocus'}) with  $\a=0$, i.e. when the potential $u(x)$ is
indeed bispectral, and we might expect a similar restriction from
implementing the higher Virasoro flows \cite{35}.

\section {Perspective}
We have already stressed some of the relevant issues we would like
to pursue in the next future. Now we are interested in delineating
and summarise the main perspective. Essentially, we want to obtain
information about the geometry of the locus (\ref{dglocus'}) and
any possible algebraic structure connected to it. Of course, an
important analysis tool should be the Virasoro algebra of flows,
since the KdV hierarchy played a crucial r\^ole in the study of
the Airault-McKean-Moser locus \cite{AMM}. Besides, the centerless
character of the algebra can be emended looking at the action on
the $\tau$-function, in terms of which the potential reads
$u=-2\p^2_x\ln\tau$ (cfr. e.g. \cite{GO} and the second of
\cite{CDD} about the KP hierarchy). In this scenario the relation
with a possible {\it generalised} bispectrality is still
mysterious, though most intriguing.

In the end, we would like to emphasise the importance of
considering non-rational forms of the potential $u(x)$, like for
instance the soliton case (for which the action of the Virasoro
algebra has been already described in \cite{FS4}).

\bigskip

{\bf Acknowledgements} - The author thanks: Leverhulme Trust
(grant F/00224/G) for a fellowship, E. Corrigan and J. Suzuki for
interesting discussions, the Asian Pacific Center for Theoretical
Physics for an invitation during which part of this work has been
performed, the EC FP5 Network EUCLID (contract number
HPRN-CT-2002-00325) for partial financial support.

\end{document}